\documentclass[a4paper,11pt]{article}
\pdfoutput=1 % if your are submitting a pdflatex (i.e. if you have
\usepackage{jinstpub} % for details on the use of the package, please
\title{\boldmath The CMS Precision Proton Spectrometer timing system: performance in Run 2, future upgrades and sensor radiation hardness studies}

\author[a]{E. Bossini,\note{Corresponding author.}}

\affiliation[a]{CERN, Esp. des Particules 1, 1211 Meyrin, Switzerland}

\emailAdd{edoardo.bossini@cern.ch}

\abstract{Central exclusive processes can be studied in CMS by combining the information of the central detector with the Precision Proton Spectrometer (PPS). PPS detectors, placed symmetrically at more than 200 m from the interaction point, can detect the scattered protons that survive the interaction. PPS has taken data at high luminosity while fully integrated in the CMS experiment. The total amount of collected data corresponds to more than 100 fb$^{-1}$ during the LHC Run 2. PPS consists of 3D silicon tracking stations as well as timing detectors that measure both the position and direction of protons and their time-of-flight with high precision. The detectors are hosted in special movable vacuum chambers, the Roman Pots, which are placed in the primary vacuum of the LHC beam pipe. The sensors reach a distance of few mm from the beam. Detectors have to operate in vacuum and must be able to sustain highly non-uniform irradiation: sensors used in Run 2 have accumulated an integrated dose with a local peak of $\sim 5 \cdot 10^{15}$ protons/cm$^2$. The timing system is made with high purity scCVD diamond sensors. A new architecture with two diamond crystals read out in parallel by the same electronic channel has been used to enhance the detector performance. In this paper, after a general overview of the PPS detector, we describe the timing system in detail. The sensor and the dedicated amplification chain are described, together with the signal digitization technique. Performance of the detector in Run 2 is reported. Recently the sensors used in Run 2 have been tested for efficiency and timing performance in a dedicated test beam at DESY. Preliminary results on radiation damage are reported. Important upgrades of the timing system are ongoing for the LHC Run 3, with the goal of reaching an ultimate timing resolution better than 30 ps; they are also discussed here.}


\collaboration[c]{on behalf of CMS and TOTEM collaborations}

\proceeding{15$^{\text{th}}$ Topical Seminar on Innovative Particle and Radiation Detectors
(IPRD19)\\
  14-17 October 2019\\
  Siena, Italy}

\begin{document}
\maketitle
\flushbottom

\section{Introduction}
\label{sec:intro}

\subsection{The PPS detector}
\label{sec:pps_det}
The Precision Proton Spectrometer (PPS) is one of the CMS\cite{CMS_std} subdetectors, born from a collaboration between the CMS and TOTEM\cite{TOTEM_tdr} experiments (and hence previously named CT-PPS\cite{CTPPS}). The detector has been designed to extend the physics program of CMS to Central Exclusive Processes (CEP) in the standard high-luminosity fills at the Large Hadron Collider (LHC). In CEP processes ($pp\rightarrow p\oplus X \oplus p$) both protons remain intact after the interaction and a pseudorapidity\footnote{Pseudorapity is defined as $\eta=-\ln\tan\frac{\theta}{2}$ where $\theta$ is the angle between the particle trajectory and the z-axis.}-isolated system X is generated in the central region. CEP are indeed characterized by the presence of large, non exponentially suppressed, pseudorapidity gaps (denoted with $\oplus$) .
Such processes can be driven by a photon-photon interaction or double Pomeron exchange.
PPS is designed to tag and measure the surviving protons kinematics. Reconstruction of mass and momentum of the central system X can be carried out from the proton information (i.e. $M_X = \sqrt{\xi_1 \xi_2 s}$, where $\xi$ represent the proton fractional momentum loss measured by PPS) and compared with the central CMS measurements for a strong background rejection. First physics results using the PPS spectrometer have been recently published\cite{PPS_dilepton}.

\begin{figure}[htbp]
\centering % \begin{center}/\end{center} takes some additional vertical space
\includegraphics[scale=0.51]{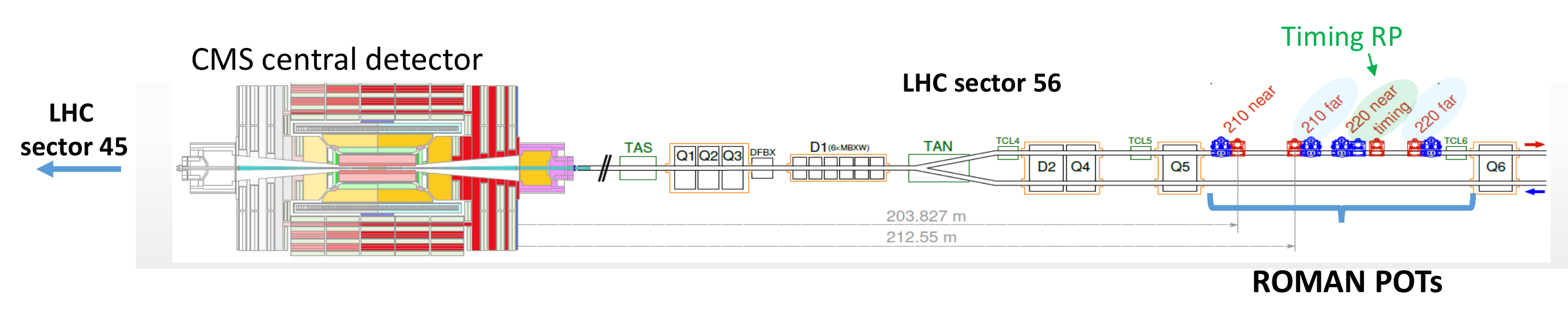}
\caption{\label{fig:PPS_layout} Layout of the CMS detector in IP5. One of the two symmetric sectors of PPS is represented in detail. RPs are located between 210 and 220 m from the IP, with the timing station in between the two tracking stations. The RP system includes stations used only by TOTEM and not part of PPS.}
\end{figure}

The detection of very forward protons is performed in movable beam insertions called Roman Pots (RP), symmetrically located at more
than 200 m from the LHC Interaction Point 5 (IP5) (Fig.\ref{fig:PPS_layout}).
The RP is a secondary vacuum vessel, hosting some type of detector, which can be
moved into the primary vacuum of the machine through vacuum bellows. The
detector can thus approach the beam down to few millimeters, which allows to
detect protons scattered down to few microradians.
Three RP units in each sector (one sector is one side of the
detector w.r.t. the IP) were used by PPS during LHC Run 2 (2016-2018), two for tracking and one for timing in between them. The mechanics and cooling have been adapted from those of the TOTEM RP. Using two RPs separated by a few meters and equipped with
tracker detectors it is possible to tag the proton and reconstruct its local position and angle. Once the position and angle of the proton track at the RP location
are known, it is possible to compute the kinematics of the proton at the interaction point by inverting the transport matrix,
which describes the propagation of the proton through the magnetic fields of the machine from the IP to the RP location. During the LHC Run 2 the PPS tracking system made use of edgeless silicon strip detectors from the TOTEM experiment\cite{TOTEM_tdr} and the newly designed monolithic 3D silicon pixel detectors\cite{PPS_pixel}. The TOTEM silicon strip detectors, designed (and used by TOTEM) for low luminosity, have no multi-tracking capability and limited radiation hardness. The pixel tracking system has been developed to overcome such limitations. Only the pixel detectors will be used for tracking during the LHC Run 3 (starting in 2021).

\subsection{Timing system requirements}
\label{sec:ts_requirements}
The average number of interactions per proton bunch crossing at LHC, also referred as pile-up, was 35 in 2018, with all vertices confined in a volume of few centimeters length (bunch longitudinal dimension is $\sigma_Z\sim7.5$ cm). Due to their location the tracking RPs are not able to reconstruct the primary interaction vertex. The solution relies on the measurement of the proton times of flight (TOF) in the two sectors. By measuring the difference $\Delta t$ of the proton TOFs is indeed possible to reconstruct the longitudinal vertex position as $Z_{pp}=c\Delta t/2$ and correlate the two protons to one of the vertices reconstructed by the central CMS instrumentation. Moreover it is possible to check if the two protons are coming from the same vertex or if they are generated by pile-up background. Detailed simulations reported in \cite{CTPPS} show the capability of the system to disentangle the pile-up with a resolution in the range 10-30 ps and a pile-up of 50.
The requirements for the timing sensor can be summarized as follow:
\begin{itemize}
  \item Station resolution in the range 10-30 ps and high efficiency in detection of 6-7 TeV protons (which can be considered as minimum ionizing particles);
  \item high radiation hardness with non-uniform irradiation field;
  \item low density and thickness of the sensor to reduce material budget and allow more planes to be fitted in the same RP;
  \item segmentation of the detector with different pad geometries, to reduce multiple hit probability in the same pad during the same bunch crossing and to keep the number of channels at minimum. Moreover it must be able to sustain particle rate up to few MHz/channel;
  \item must be operated in vacuum.
\end{itemize}

\begin{figure}[htbp]
\centering % \begin{center}/\end{center} takes some additional vertical space
\includegraphics[scale=0.71]{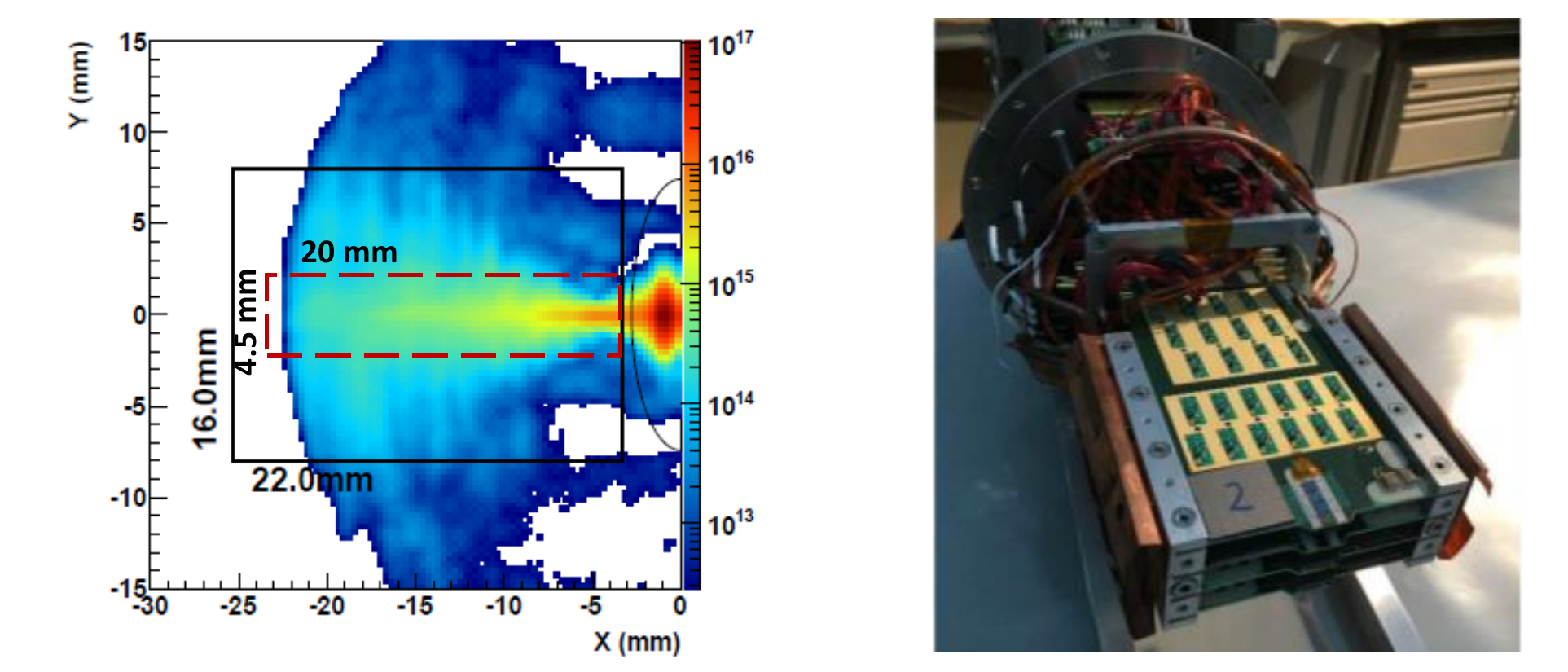}
\caption{\label{fig:Detector_intro} On the left the simulated proton fluence (proton/cm$^2$) in  the PPS sensor area for 100 fb$^{-1}$ of integrated luminosity delivered by the LHC\cite{CTPPS}. The red dashed box represent the area covered by the timing sensors, while the black solid line is the area covered by the pixel tracker. On the right a complete timing detector package, hosted in one RP, composed of 4 detection planes.}
\end{figure}

The requirement on radiation hardness is particulary tight, since the sensors have to sustain a highly non uniform irradiation (see Fig. \ref{fig:Detector_intro} left), with a peak of  $\sim5\cdot10^{15}$ protons/cm$^2$ in the near beam region for an integrated LHC luminosity of 100 fb$^{-1}$ (which represent the order of magnitude delivered by LHC with RP inserted during Run 2 and foreseen for Run 3).
The technology chosen by PPS is based on ultrapure single crystal chemical vapour deposition (scCVD) diamonds. Since 4 detector planes can be hosted in a single RP (see Fig. \ref{fig:Detector_intro} right), the timing requirements on a single detection plane are less stringent. During Run 2 one plane based on Ultrafast Silicon Detectors\cite{UFSD} has also been used for R\&D purpose. Diamond technology has been confirmed for the Run 3.

\section{The PPS timing system}
\label{sec:pps_timing_system}

\subsection{Sensor and front-end amplification}
\label{sec:ts_frontend}

\begin{figure}[htbp]
\centering % \begin{center}/\end{center} takes some additional vertical space
\includegraphics[scale=0.51]{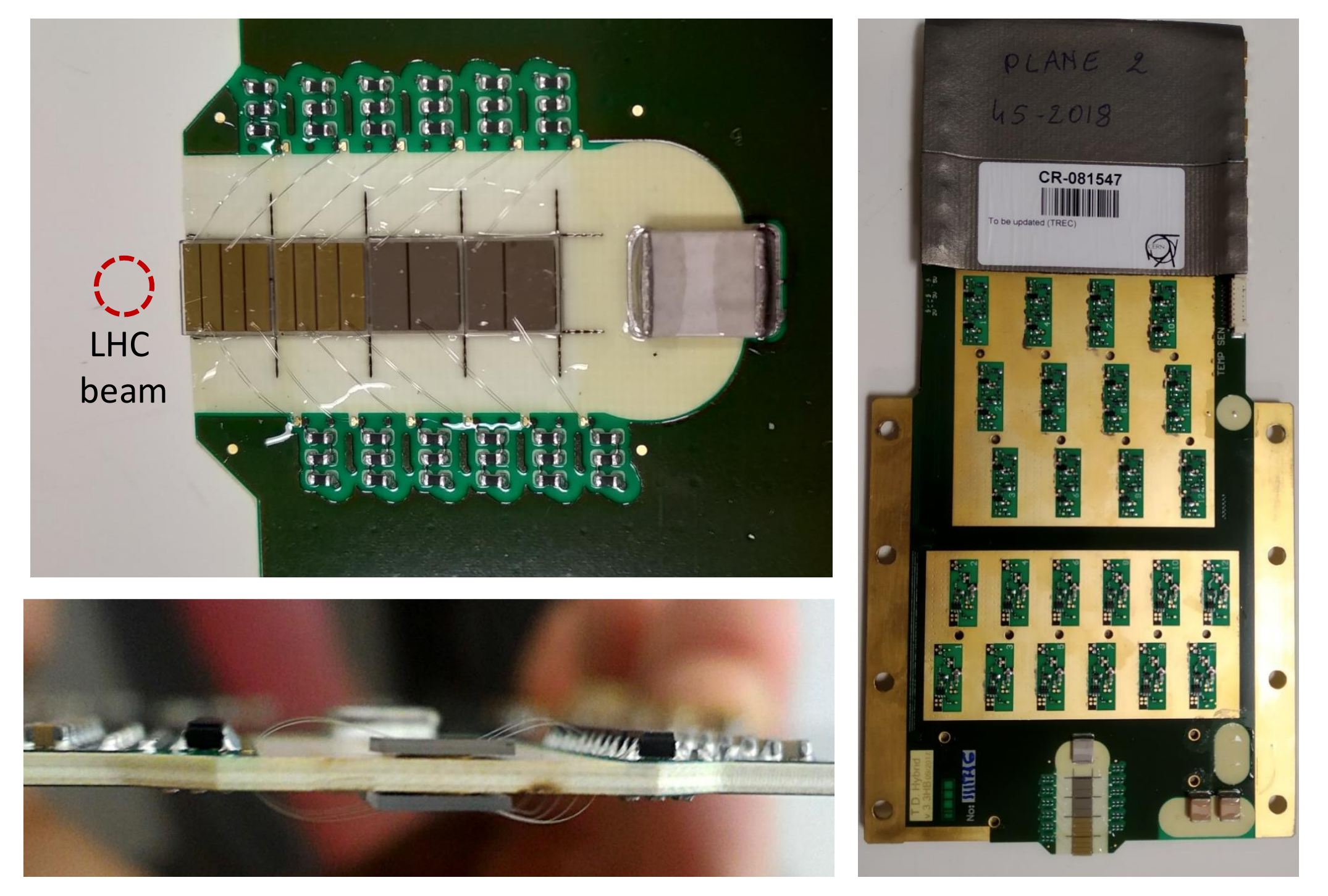}
\caption{\label{fig:diamond sensor} Pictures of the PPS timing sensors. On top left closeup view of the crystals: sensor segmentation in strips and the bonding wires are visible. On the right the full hybrid board hosting sensors, amplification channels and HV distribution. On the bottom left a lateral view of the a double diamond sensor, with crystals glued on both sides of the hybrid board.}
\end{figure}

The PPS diamond sensors, developed by the TOTEM and CMS Collaborations, are made of scCVD crystals with a surface of 4.5x4.5 mm$^2$ and a thickness of 500 $\mu$m, with a total active surface coverage $\sim$20x4.5 mm$^2$ (Fig. \ref{fig:diamond sensor}). The crystal segmentation is carried out in the metallization phase on the top face of the diamond, performed at GSI (Darmstadt, Germany) and PRISM (Princeton, USA). The bottom side of the crystal, where HV is applied, is instead metallized with a single pad. Different metallization procedures have been used, tested to be equivalent in terms of performance during test beams. Strips on the same crystals are separated by 100 $\mu$m and a clearance area of 150 $\mu$m is taken from the crystal edges.

The diamonds are glued to a hybrid board (Fig. \ref{fig:diamond sensor}) with 12 discrete amplification channels, designed and optimized for diamond signals. The amplification is performed in three stages. Signals from crystals can be modeled by a triangular current pulse with a maximum amplitude of $\sim$1$\mu$A, a rising edge of few ps and a falling edge of few ns. Intrinsic noise of the crystal is very low, below the nA level, and care must hence be placed in the pre-amplification stage, the noise being dominated by its input stage. We use  a transconductance amplifier (BFP840 SiGe BJT) in a common emitter configuration. Sensor strips are directly connected to the pre-amplifier input to reduce the  parasitic capacitance (estimated $\sim0.2$ pF with 0.25$\mu$m bonding wire diameter). With this architecture the input capacitance is dominated by the strip capacitance (0.2 to 2 pF depending on its size).
The signal from the first stage is fed to the second stage built around a monolithic microwave integrated circuit (Avago ABA-53563, which will be replaced with GALI-39+ for Run 3). The final amplification stage is designed to shape the signal, and is composed of two wide-band BJT transistors NPX BFG425W. Since the detector has to operate in vacuum with a nominal bias voltage $\sim$500 V, a special coating is applied to sensitive areas to reduce discharge probability. More details about sensors, amplification and test beam performance can be found in \cite{totem_diamond_article}.

\begin{figure}[htbp]
\centering % \begin{center}/\end{center} takes some additional vertical space
\includegraphics[scale=0.62]{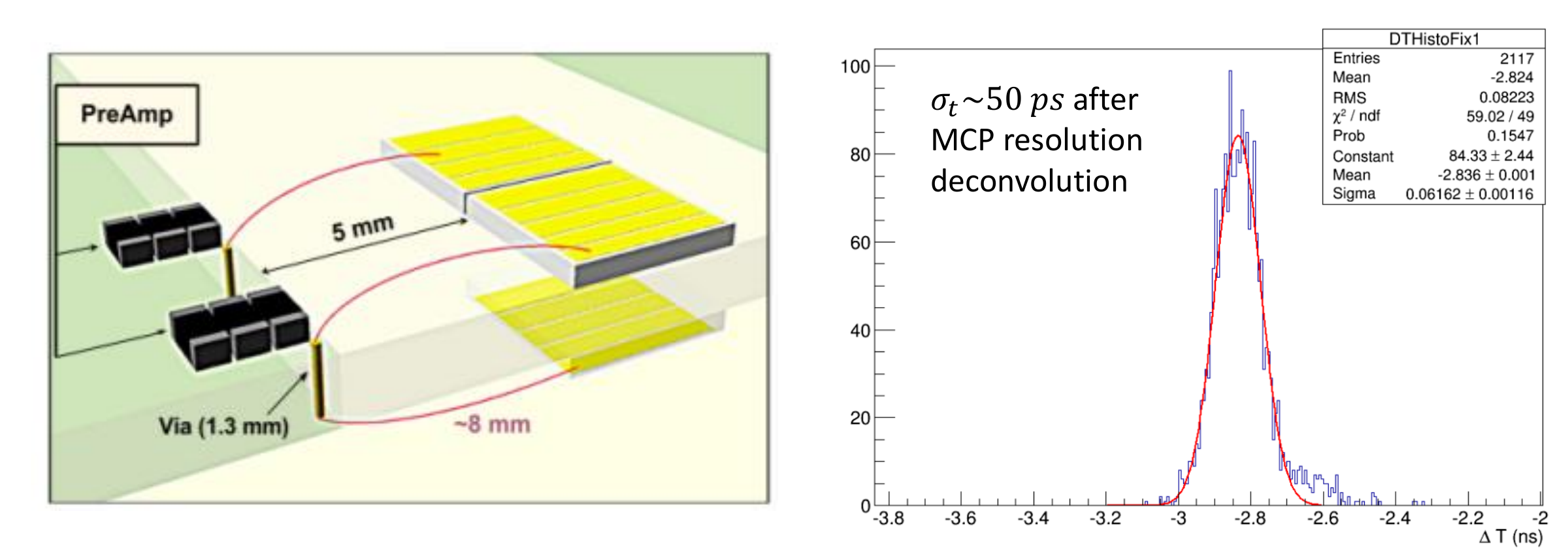}
\caption{\label{fig:DD_perf} On the left, a scheme of the DD connection with the corresponding strips connected to the same amplification input. On the right, time difference distribution between a DD plane and a reference Micro Channel Plate (MCP). Details of the measurement can be found in \cite{DD_article}.}
\end{figure}

A first version of the sensors (referred to as Single Diamond architecture, SD) was used for the full Run 2, while a new architecture (Double Diamond, DD) was developed for the 2018 data taking and some SD planes previously used were replaced.
DD architecture (Fig. \ref{fig:diamond sensor}) has been developed to achieve better time performance\cite{DD_article} on a single detector plane. In this design diamond crystals with the same metallization geometry are glued on both sides of the hybrid board. Signals from corresponding pads are connected to the same amplification channel, as seen schematically in Fig. \ref{fig:DD_perf}. With this architecture it is possible to double the signal, while keeping the noise unchanged (dominated by the pre-amplification input). This goes at the expense of a higher sensor capacitance and the necessity of a very precise alignment of the crystals. The loss of resolution due to higher sensor capacitance is however much lower w.r.t. the gain introduced by the higher signal amplitude. In test beam, operating the sensor in nominal condition and using an Agilent DSO9254A oscilloscope (8 bits, 20 GSa/s) a time resolution of $\sim$50 ps with a single plane of DD has been obtained (Fig. \ref{fig:DD_perf}). This results in a better time resolution by a factor $\sim$1.7 w.r.t. the SD architecture, which was measured in test beam in the range 80-100 ps, depending on the strip size. The increase in performance is superior to that achievable by simply doubling the number of detection planes. Moreover such improvement is achieved while maintaining the same number of hybrid boards and the same number of channels to be read out and controlled.

\subsection{Digitization and readout}
\label{sec:ts_digi}

\begin{figure}[htbp]
\centering % \begin{center}/\end{center} takes some additional vertical space
\includegraphics[scale=0.55]{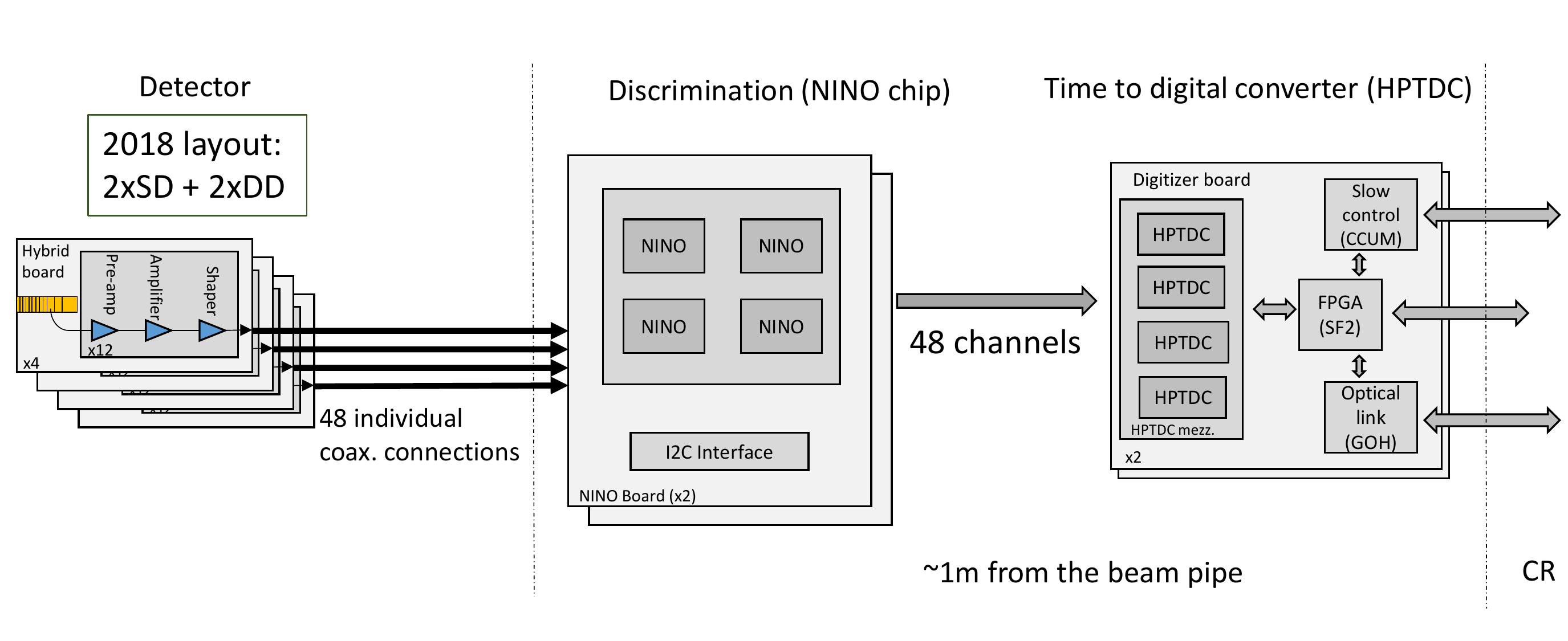}
\caption{\label{fig:timing_layout} Scheme of one sector of the PPS timing system in 2018.}
\end{figure}

In Fig. \ref{fig:timing_layout} we report the scheme of one sector of the PPS timing system, as used in 2018, with 2 SD and 2 DD planes in each RP.
When performing time measurements care must be put not only in the intrinsic time resolution of the sensor, but also in the uncertainties introduced during the signal digitization process. A widely used approach is to use a discriminator coupled to a Time to Digital Converter (TDC). In this case, if a fixed threshold is used, an error (often referred to as \emph{time walk}) is introduced. Indeed, due to the statistical fluctuations of the energy release in a thin detector, the same particle passing through two consecutive sensors will release a different energy. The output signal of the sensor where more energy was released will thus go above threshold before the other. The time walk can be the dominant source of uncertainty if not properly corrected, especially when large fluctuations of energy release are foreseen. The most common technique to mitigate the problem is represented by the Constant Fraction Discriminator (CFD), where the threshold of each signal is put at a certain percentage of its maximum. Another way to correct the time walk effect is to apply a time correction using the Time Over Threshold (TOT) measurement. TOT is the measurement of the time during which the signal remains above the threshold, and is correlated to the collected charge. For the PPS readout we have pursued this latter strategy.

For the discrimination we selected the NINO chip\cite{nino}, an 8-channel ultra-fast
low-power differential amplifier and fixed threshold discriminator, with an input range 0.01-2 pC.
The discrimination
threshold can be adjusted in the range 10-100 fC through the differential voltage
applied to dedicated pins of the chip.
In addition to the signal leading edge measurement the chip can encode the input charge
Q collected from the detector in the output duration, $W=K+f(Q)$, which can be offline used for TOT corrections. A constant stretch time K ($\sim$10 ns) is added to cope with the next stage of the readout, the widely used HPTDC\cite{hptdc}.
The HPTDC can be configured to operate with a binning of 25 ps, which leads to a nominal resolution $\sim$7 ps. The HPTDC is also able to measure both the leading and trailing edges of the signal, provided that they are separated by at least 5 ns. Two different boards, one for the NINO and one for the HPTDC chips have been developed and are located in the detector area, around 1 m above the beam pipe. For such reason all related electronics has been designed as radiation tolerant (including FPGAs).

A different approach to perform timing measurements is instead based on a fast sampling of the signal in order to use sophisticated offline reconstruction algorithms and obtain the best performance. Drawback of this method is that the sampler must have negligible dead time and be able to sustain a high input rate. Moreover this approach produces a larger data stream with respect to the TDC, where a maximum of one or two time measurements are performed on each waveform. This strategy has been exploited in the TOTEM timing system, used in a special CMS-TOTEM common data taking in 2018 involving the vertical RPs\cite{twepp}. We return to this subject in section \ref{sec:perspectives}.

\section{Run 2: calibration and performance}
\begin{figure}[htbp]
\centering % \begin{center}/\end{center} takes some additional vertical space
\includegraphics[scale=0.52]{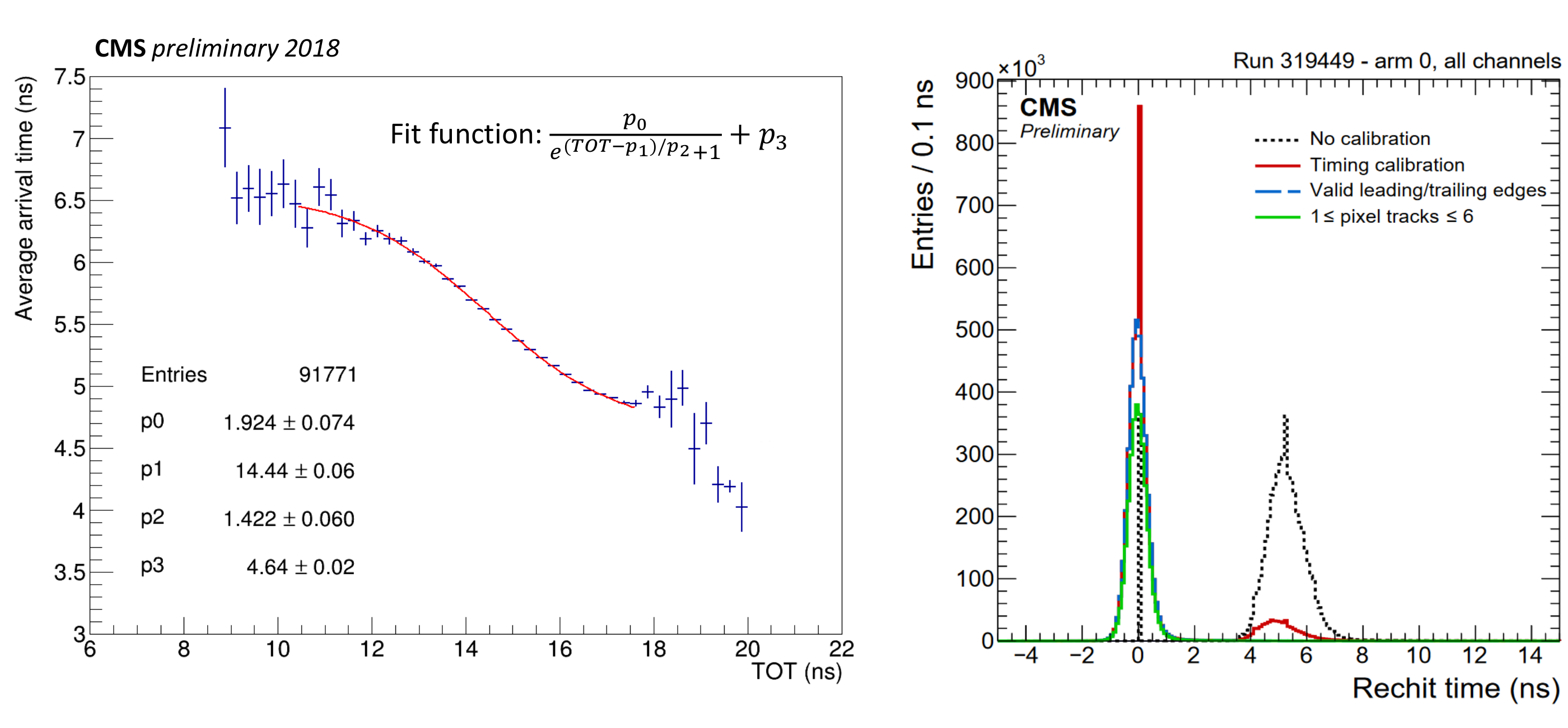}
\caption{\label{fig:calibration} On the left an example of a calibration curve for one timing channel. On the right the effect of the calibration. The black dashed curve represents the distribution of the raw timing measurements from all channels of one sector, the other curves represent the distributions after correction, applying different quality cuts. Plots from \cite{dps_note}.}
\end{figure}

The timing system was first installed in late 2016. The best performances and system stability were reached in 2018, after LHC Technical Stop 1 (TS1) in July 2018. Offline calibration  has been performed for the 2017 and 2018 data and included in the most recent CMS data reconstruction. The calibration procedure is divided in two steps. First the profile distribution of the average arrival time of the protons vs the signal TOT is extracted for each individual channel (Fig.\ref{fig:calibration}).
\label{sec:performance}
%The average arrival time is here computed w.r.t. the closest LHC clock cycle, meaning that an overall offset for all the channel in the RP is subtracted without any impact on the calibration.
As expected, signals with higher TOT have lower average arrival time, due to the time walk effect mentioned before. The curve can be fitted and the data corrected. With this procedure it is also possible to align all channels in time. In Fig. \ref{fig:calibration} (right) it is possible to appreciate the effect of the calibration on all the channels of one sector. The width of the corrected distribution is dominated by the bunch length, since no time difference between the two sectors is computed at this stage.
The second step of the calibration consists of an iterative procedure to compute the resolution of each channel. This is indeed extremely important since each plane has different performance and planes with different technologies are hosted in the same RP. The resolutions obtained are then used to compute a weighted average of the time measurements performed on the proton, computing its TOF with the best accuracy. Moreover an expected resolution can be assigned to each particle depending on the channels that contributed to the time measurement. Such information is stored together with the proton time in the reconstructed data for later analysis.

\begin{figure}[htbp]
\centering % \begin{center}/\end{center} takes some additional vertical space
\includegraphics[scale=0.48]{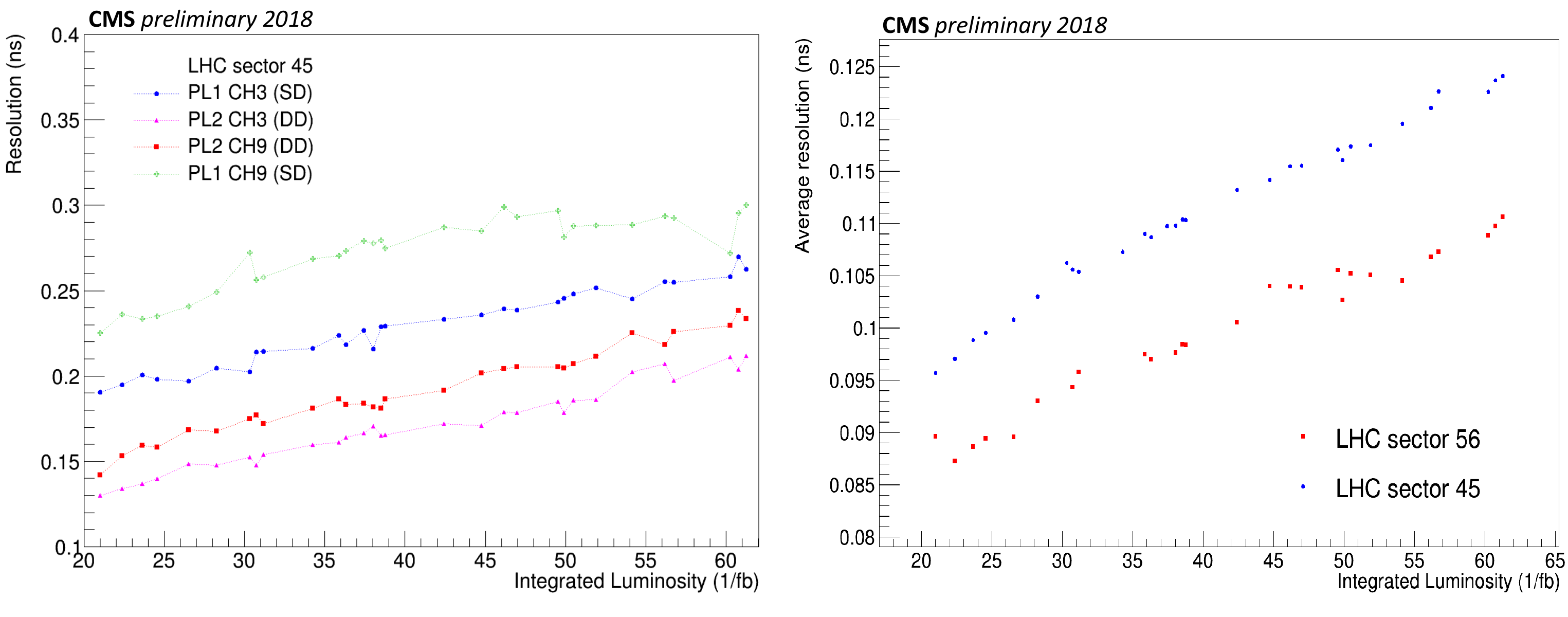}
\caption{\label{fig:performance} On the left evolution of 2 SD and 2 DD channels with almost equal radiation exposure. On the right the evolution of time resolution of each sector. In both cases data start from 2018 TS1, with sensors having already collected $\sim$20 fb$^{-1}$. Plots from \cite{dps_note}.}
\end{figure}

Figure \ref{fig:performance} (left) shows the resolution for some channels as a function of the integrated luminosity delivered by the LHC with the RPs inserted (which is in turn linearly correlated with the sensor irradiation).
As expected from the test beam, the resolution of SD is a factor $\sim$1.7 worse w.r.t. DD channels.
On the right side the average proton time resolution for each station is shown. The values represent the convolution of station resolution, digitization effects and calibration quality.  Other effects will have to be taken into account, and corrected for, when performing correlation between the two sectors (clock drift in the transmission line to the RP location, beam phase w.r.t. LHC clock, etc). Such fine tuning, which needs to be combined with some physics analyses, is currently ongoing and will lead to the final measurement of the $Z_{pp}$  vertex resolution. The proton time resolutions reported here have been obtained by using the tracking stations to select single track events and requiring at least one hit on each timing layer; efficiency studies are ongoing. We obtained an overall station performance in the range 90-120 ps depending on the sector and on the accumulated luminosity. The resolutions obtained are worse than the one expected from previous test beam results due to three main limiting factors encountered during Run 2. Radio frequency oscillations were identified in the timing RP after installation, forcing a reduction of the front-end LV in order to keep the amplifiers in a stable regime. Moreover beam induced discharges did not allow us to operate the sensors with nominal HV (350-400 V were used instead of 500 V). The effect of reduced voltages has been estimated in the range 30-40\% during the test beam described in section \ref{sec:radiation}. Finally the coupling of the NINO with the sensor was not fully optimized, leading to a $\sim$30\% degradation of detector time resolution. Work is ongoing to remove such limitations for the upcoming Run 3 (see sec. \ref{sec:perspectives}).
The performance achieved for Run 2 allows to use the timing information in physics analyses to perform an important background reduction.
\section{Radiation effects}
\label{sec:radiation}
\begin{figure}[htbp]
\centering % \begin{center}/\end{center} takes some additional vertical space
\includegraphics[scale=0.71]{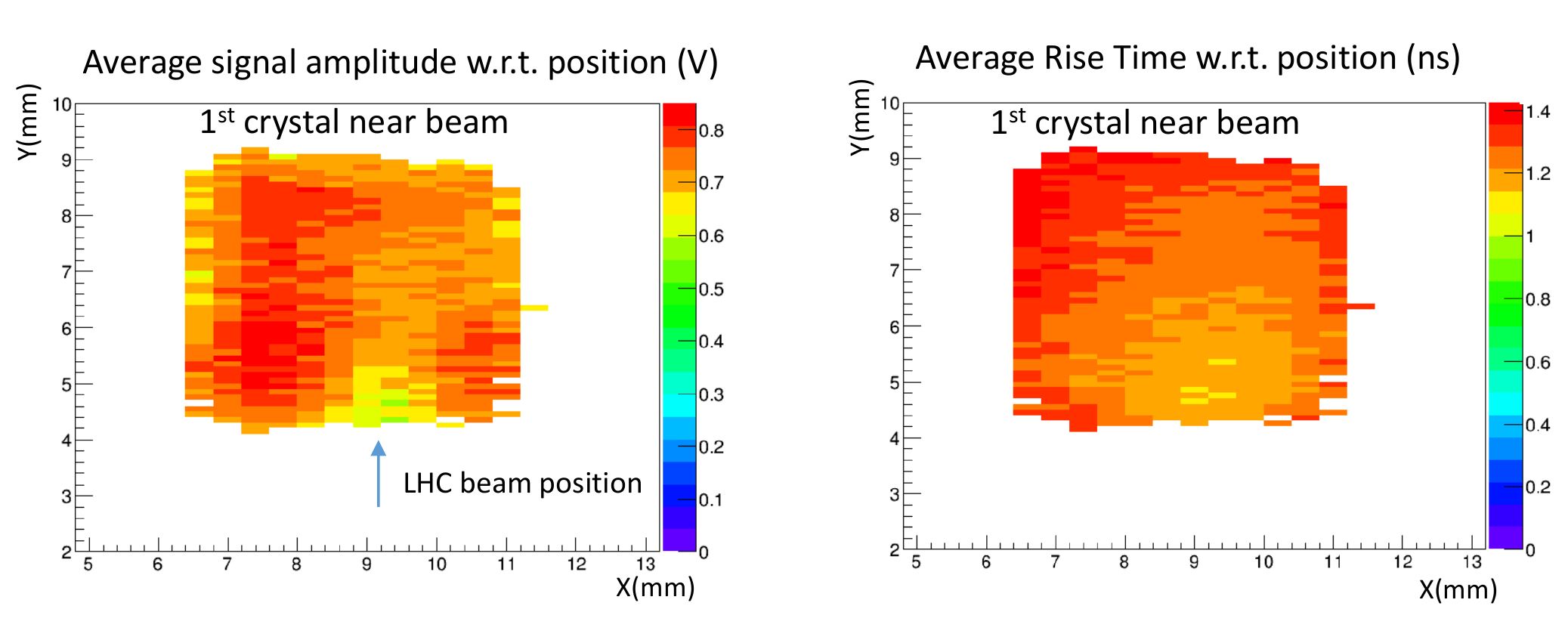}
\caption{\label{fig:test_beam} Average signal amplitude (left) and rise time (right) as a function of the position for the crystal closest to the LHC beam.}
\end{figure}

Concerning radiation damage, we have identified two main types of effect. The first one, well visible in the plots discussed above, is a generalized performance loss. The degradation does not show a clear correlation with the position of the pad and hence with its irradiation. The reduction of resolution is in the 20-50\% range for the full 2018 data taking. This could be due to radiation damage to the pre-amplification stage, which is not protected by the collimators as the rest of the electronics and is thus immersed in the beam halo. In test beam it has been measured that increasing the pre-amplifier voltage can compensate  for this effect, but unfortunately remote control of pre-amplifier LV was not possible during Run 2.
The second type of damage is instead well localized in a small area of the detector ($\sim$1 mm$^2$) close to the beam. The damage has been studied in detail during a test beam campaign carried out at DESY (Hamburg, Germany) during May 2019 with a 4.8 GeV electron beam. The timing layers used in 2018 have been dismounted and brought to the DESY test beam line T24, where a tracker of the EUDET series\cite{eudet} is available. The tracker is composed of six planes, three upstream and three downstream the device under test. In order to reduce material budged (and thus electron multiple scattering) only two timing layers at the time were tested, with almost nominal LV and HV settings.
With the tracker it is possible to reconstruct with a precision $\sim$ 100 $\mu$m (mainly limited by the multiple scattering effect) the position of the electron in the diamond planes.
Figure \ref{fig:test_beam} reports the average signal amplitude and rise time measured on the first crystal (the one closest to the beam).
Signals from diamonds were acquired with the SAMPIC\cite{sampic} chip (a fast sampler operated at 11 bit resolution and 6.4 GSa/s sampling frequency) and merged offline with the tracker data. Event synchronization was provided by the telescope trigger logic unit, which was delivering the event number to both tracker and SAMPIC.
The area close to the beam shows a reduced average signal amplitude, partially compensated by a faster signal rising edge. The efficiency in this situation has been measured still >95\%, also in the most irradiated region. The observed radiation damage can affect the bulk of the crystal or the metallization surface (or both). In the next months we will unglue, etch and re-metallize the crystals used in Run 2 and a new test campaign will lead to a better understanding of the phenomena. During the test beam also timing performance have been measured and the results will be soon available.

\section{Run 3 upgrades}
\label{sec:perspectives}

An important upgrade program is ongoing for Run 3, with the goal of reaching a resolution better then 30 ps on each sector.
A new hybrid board has been designed and is currently under test. Effort has been put to increase the amplification stability and radio-frequency shielding, improving at the same time the high voltage isolation. During Run 2 we have understood the importance of being able to remotely control the LVs provided to each stage of the amplification chain. In Run 3, remote LV control will be implemented, allowing us to operate the sensor always in the best possible condition and to perform a compensation of the radiation damage.

A new discriminator board is also under production. The new board, still based on the NINO chip, will be used to test different connection schemes with the hybrid board, to reduce the timing degradation at the digitization phase.

A major upgrade will consist in the implementation of a parallel independent readout based on the SAMPIC chip, a fast sampler. This readout, already used for TOTEM timing\cite{twepp}, is indeed already integrated in CMS. It will be operated only in low intensity fills or on a subset of bunches due to rate limitations. Nevertheless it will provide valuable information to calibrate and monitor the sensor performance by having a sampling of their output analog signals.

Finally PPS will build and install a second timing RP on each sector, and all stations will be equipped with 4 planes of DD layers. This will provide a total of 8 DD layers per sector, instead of 2 SD and 2 DD layers as in RUN 2. The increase in the number of layers will allow to reach a resolution of 30 ps, with a safety margin.

\section{Conclusions}
\label{sec:conclusions}

TOTEM and CMS have developed new timing detectors based on scCVD diamonds. The new Double Diamond architecture described here brought a significant improvement  (factor 1.7) over the Single Diamond architecture, maintaining the same number of layers and readout channels. A single layer has proved to reach a resolution $\sim$ 50 ps when operated under nominal conditions and read out with an oscilloscope or a sampler chip. RP stations based on this technology have been built and operated during LHC Run 2 to provide timing information on the protons tagged by the PPS subdetector. To cope with the data size and rate requirements, the readout is based on a fast discriminator (NINO), coupled to the high precision HPTDC.
Performance results in Run 2 allow to use timing information to reduce background in CEP processes; the longitudinal vertex reconstruction resolution is under study.
The collaboration has performed important studies on the effect of radiation on crystals and electronics. Further studies will be carried out in the near future.
Results will be extremely relevant due to the high level and non uniformity of the particle flux.
Important upgrades are ongoing for Run 3, with the ultimate goal of a timing resolution  better then 30 ps in each sector within reach.

\acknowledgments

The test beam measurements on irradiated samples reported here were performed at the Test Beam Facility of DESY, Hamburg (Germany), a member of the Helmholtz Association (HGF).

\end{document}